\documentclass[aps,pra,twocolumn,showpacs,superscriptaddress,10pt]{revtex4-1}
\usepackage[colorlinks ,linkcolor=blue,anchorcolor=blue,citecolor=blue,urlcolor=blue]{hyperref}
\usepackage{amsmath,amssymb}
\usepackage{bm}
\usepackage{graphicx}
\usepackage{epstopdf}
\usepackage{color}

\usepackage{float}
\usepackage{color}
\usepackage[T1]{fontenc}
\usepackage{tgtermes}

\begin{document}

\title{Quantum Monte Carlo simulations of thermodynamic properties of attractive SU($3$) Dirac fermions}
\author{Xiang Li}
\affiliation{School of Physics and Technology, Wuhan University, Wuhan
430072, China}
\author{Han Xu}
\affiliation{School of Physics and Technology, Wuhan University, Wuhan
430072, China}
\affiliation{Department of Physics, City University of Hong Kong, Tat Chee Avenue, Kowloon, Hong Kong SAR, China, and City University of Hong Kong Shenzhen Research Institute, Shenzhen, Guangdong 518057, China}
\author{Yu Wang}
\email{yu.wang@whu.edu.cn}
\affiliation{School of Physics and Technology, Wuhan University, Wuhan
430072, China}

\begin{abstract}
  We employ the determinant quantum Monte Carlo method to study the finite-temperature properties of the half-filled attractive SU($3$) Hubbard model on a honeycomb lattice. We calculate the phase diagram in which the phase boundary separates the disordered phase and the charge-density-wave (CDW) phase and the transition temperature $T_{\text{tr}}(|U|)$ varies non-monotonically with attractive Hubbard interaction $|U|$. As the Hubbard $|U|$ increases at constant temperature $T<\text{max}(T_{\text{tr}}(|U|))$, the system first undergoes a transition from thermal Dirac semimetal phase to CDW phase, and eventually the CDW state is thermally melted at a strong Hubbard $|U|$ where the system enters a trion liquid phase. In between the two transition points the non-monotonic $|U|$ dependence of CDW order strength is strikingly different from the zero-temperature monotonic behavior. In the trion CDW state where off-site trions arise from quantum fluctuations (a fermion inside an on-site trion hops to a nearest-neighbor site), the simulated triple occupancy at constant Hubbard $|U|$ surprisingly increases with temperature, implying that the formation of off-site trions is suppressed by the thermal delocalization of on-site trions. We have also calculated the entropy-temperature relations for various attractive Hubbrad interactions, which exhibit the prominent characteristic of the Pomeranchuk effect. Our work has revealed that the formation of on-site and off-site trions has significant consequences for thermodynamic properties of SU(3) Dirac fermions.
\end{abstract}

\maketitle

\section{Introduction}
Ultracold fermions provide highly tunable systems for studying SU($N$) ($N>2$) physics which is common in high-energy context but rare in solids. In the alkaline-earth fermions that simply carry large hyperfine nuclear spins owing to closed-shell electronic structure, SU($2N$) symmetry arises when the interatomic scatterings are spin-independent \cite{Gorshkov2010,Wu2012,Taie2010,Desalvo2010}. In alkaline fermions such as $^{6}\text{Li}$, the SU($3$) symmetry arises when the $^{6}\text{Li}$ atoms have equal populations in the three lowest energy hyperfine states (often referred to as "colors") and the pairwise $s$-wave scattering lengths between these colors approach a common value \cite{Ottenstein2008,Huckans2009}. Unlike the classic-like large-$S$ scenario, the large number of multiple components $N$ can significantly enhance quantum fluctuations and thus induces even richer quantum phases in SU($N$) ultracold fermions compared to spin-$1/2$ electronic systems \cite{Wu2012,Wu2010}. In recent two decades the study on novel states of ultracold fermionic atoms with SU($N$) symmetry has become one of major research foci at the interdisciplinary frontiers of cold atom physics and condensed matter physics \cite{Cai2013,Wang2014,Zhou2014,Zhou2016,Zhou2017,Zhou2018,Gorelik2009,Inaba2010,Miyatake2010,Inaba2012,Inaba2013,Okanami2014,Suga2015,Yanatori2016,Yanatori2016b,Hasunuma2016}.

Among numerous SU($N$) fermionic systems, the attractive SU(3) Hubbard model has a special standing, most obviously because of formation of trions bearing resemblance to the quark matter. In a three-fermion attractive SU($3$) Hubbard model on the square lattice, the exact diagonalization calculation illustrates two configurations of trionic states \cite{Pohlmann2013}: the on-site trion composed of three fermions at one site, and the off-site trion consisting of two fermions at one site and one fermion at the nearest-neighbor site. In  many-body systems, the quantum Monte Carlo (QMC) simulations of the half-filled attractive SU(3) Hubbard model on a honeycomb lattice demonstrate the formation of a local bond state of the off-site trion in the background of the trion charge-density-wave (CDW) phase \cite{xu2019quantum}. The variational \cite{Rapp2007,Rapp2008}, self-energy functional \cite{Inaba2009,inaba2011color} and dynamical mean-field theory \cite{Titvinidze2011,Koga2017} studies find that in the attractive SU(3) Hubbard model, a phase transition between the color superfluid and the on-site trion phase occurs, which is reminiscent of the transition between the quark superfluid and the baryonic phase \cite{Fodor2002,Aoki2006,Wilczek2007}.  In an one-dimensional lattice away from half filling, the density matrix renormalization group studies show that off-site trions can develop quasi-long-range correlations, when on-site triple occupancy is prohibited in a SU(3) attractive Hubbard model \cite{Kantian2009}, or when the attractive interactions are color-dependent in a three-component Hubbard model with SU(3) symmetry breaking \cite{Capponi2009}.

It has been found that the on-site and off-site trions coexist in the ordered phase of the one-dimensional three-component Hubbard model with color-dependent interactions away from half filling \cite{Capponi2009} and the honeycomb-lattice SU(3) Hubbard model at half filling \cite{xu2019quantum}. However the research on exploring the physical effects of the interplay between on-site and off-site trions in a two-dimensional spatial model is still at the very early stage.  In this work, we propose to investigate the thermodynamic properties of the half-filled attractive SU($3$) Hubbard model on a honeycomb lattice, by performing the sign-problem-free determinant QMC (DQMC) simulations. We shall focus on the formation of trion states at finite temperatures, through which we can investigate how the on-site and off-site trions affect the thermal phase transition, triple occupancy, entropy-temperature relation and density compressibility.

The rest of this paper is organized as follows. In Sec.~\ref{sec:model}, the model Hamiltonian and parameters of DQMC simulations are introduced. In Sec.~\ref{sec:cdw-trans}, the phase diagram is obtained from DQMC simulations. In Sec.~\ref{sec:occupancy}, formation of trion states is studied via DQMC simulations of triple occupancy and density correlation function. In Sec.~\ref{sec:entropy}, the entropy-temperature relations are calculated and analyzed. Subsequently in Sec.~\ref{sec:knn}, the density compressibility is investigated. The conclusions are drawn in Sec.~\ref{sec:conclusion}.

\section{Model and method}
\label{sec:model}
The half-filled attractive SU($3$) Hubbard model on the honeycomb lattice takes the form
\begin{equation} \label{main.Eq.1}
    H = -t\sum_{\langle ij\rangle,\alpha}(c^{\dagger}_{i\alpha}c_{j\alpha}+\mathrm{H.c.}) + U\sum_{i,\alpha<\beta}{(n_{i\alpha}-\frac{1}{2})(n_{i\beta}-\frac{1}{2})},
\end{equation}
where $\langle ij\rangle$ denotes the nearest-neighbor sites; $\alpha$ and $\beta$ are the color indices running from $1$ to $3$; the nearest-neighbor hopping amplitude $t$ is set as energy unit in our simulations; $n_{i\alpha}=c^{\dagger}_{i\alpha}c_{i\alpha}$ is the particle number operator for color $\alpha$ at site $i$; $U<0$ describes the attractive Hubbard interaction. The chemical potential vanishes at half filling.

The DQMC simulation of the half-filled attractive SU(3) Hubbard model in a bipartite lattice is sign-problem-free when the Hubbard-Stratonovich (H-S) decomposition in the color-flip channel  is employed \cite{Wang2015, xu2019quantum}. We shall adopt a mathematically rigorous H-S decomposition revised from Ref.~\cite{xu2019quantum}, as presented in Appendix~\ref{decomposition}. In our DQMC simulations, the Suzuki-Trotter discretization $\Delta \tau$ is set between $\frac{1}{12}$ and $\frac{1}{8}$. The $2\times L\times L$ honeycomb lattices with $L=3,6,9,12$ are simulated under the periodic boundary condition which preserves the translational symmetry. For a typical data point, 300--500 warmup steps and 300--500 measurements are used in QMC bins. Unless specifically stated, the temperature $T$ and the Hubbard $U$ are given in the unit of $t$.

\section{The phase diagram}\label{sec:cdw-trans}

The projector QMC simulation demonstrates that the half-filled SU($3$) Hubbard model on a honeycomb lattice can undergo a continuous quantum phase transition between the semimetal and the CDW phase at the quantum critical point $U_c = -1.52$ \cite{xu2019quantum}. Since the CDW ordering on the honeycomb lattice breaks the discrete symmetry of lattice inversion, it can survive at low temperatures. In this section, we shall investigate the thermal phase transitions of attractive SU(3) Dirac fermions.

In QMC simulations, the CDW ordering can be characterized by the CDW structure factor:
\begin{equation}
  S_\text{CDW}(L,\Gamma)=\frac{1}{2L^{2}}\sum_{i,j}(-1)^{i+j}C(i,j),
\end{equation}
where the density-density correlation function $C(i,j)=\sum_{\alpha,\beta}\langle n_{i\alpha}n_{j\beta}\rangle$.
Then the CDW order parameter is defined as
\begin{equation}
  D=\lim_{L\rightarrow \infty}\sqrt{\frac{1}{2L^{2}}S_\text{CDW}(L,\Gamma)}.
\end{equation}

\begin{figure}[b]
  \centering
  \includegraphics[width=0.96\linewidth]{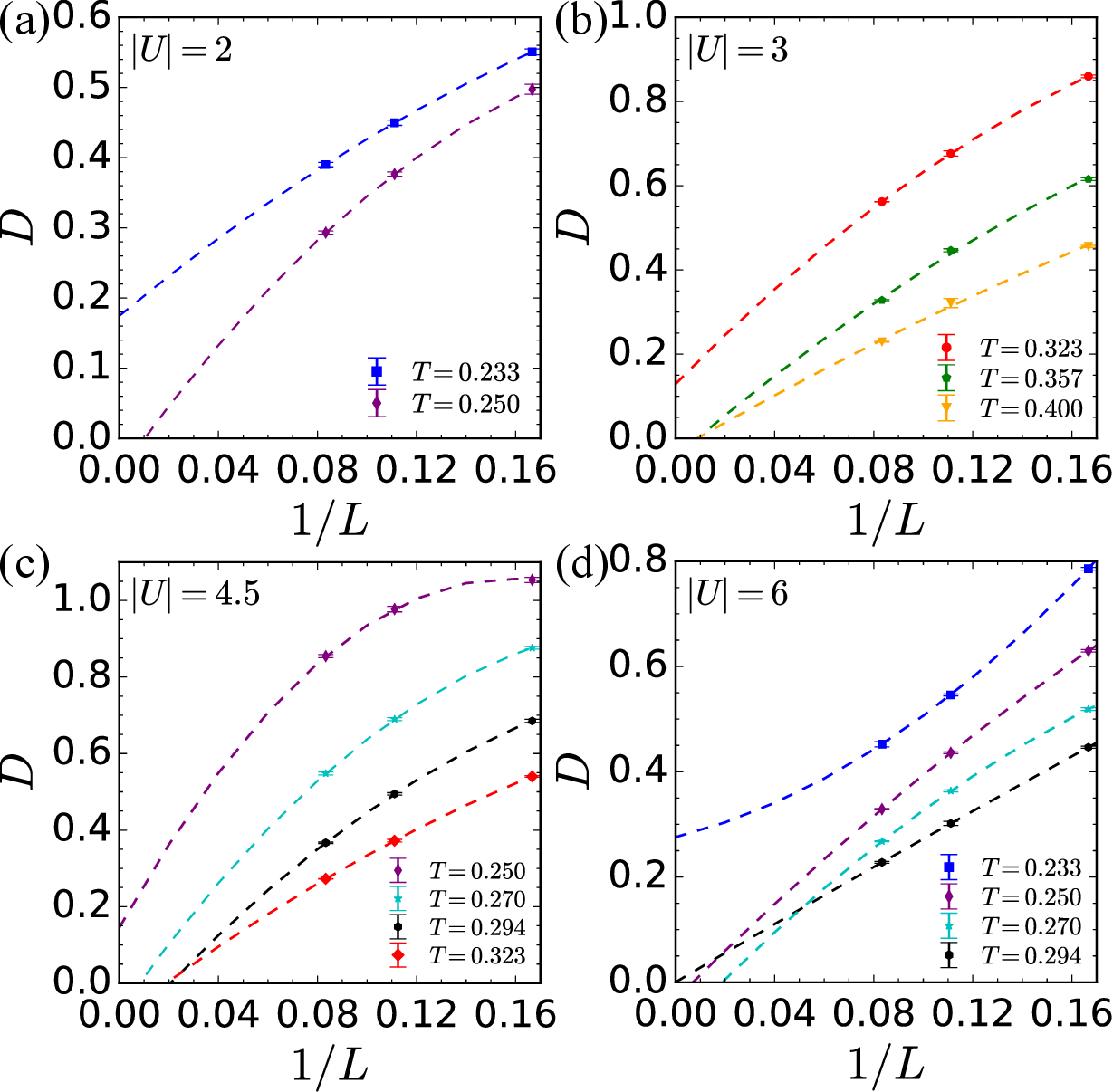}
  \caption{The finite-size extrapolations of the CDW order parameter for various Hubbard $\left|U\right|$:
(a) $\left|U\right|=2$; (b) $\left|U\right|=3$; (c) $\left|U\right|=4.5$; (d) $\left|U\right|=6$. The quadratic polynomial fitting is used.
  } \label{fig:main:D}
\end{figure}

\begin{figure}[t]
    \centering
    \includegraphics[width=0.9\linewidth]{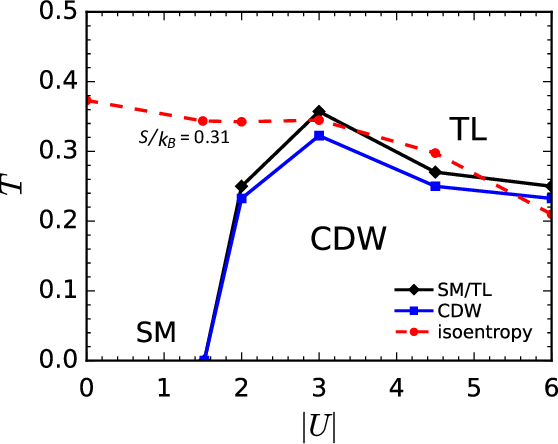}
    \caption{The finite-temperature phase diagram of the half-filled attractive SU(3) Hubbard model on a honeycomb lattice, SM -- semimetal, TL -- trion liquid, CDW -- charge density wave. The red dashed curve represents the isoentropy curve of $S/k_B=0.31$. The quantum critical point is extracted from Ref.~\cite{xu2019quantum}.
    } \label{fig:main:phasediagram}
\end{figure}

At low temperatures, the finite-size extrapolations of the CDW order parameters for various Hubbard attractions are shown in Fig.~\ref{fig:main:D}. At constant Hubbard attraction $|U|$, the CDW order develops with the decrease of temperature. The transition temperatures $T_{\text{tr}}$ of the CDW transitions for $|U|=2, 3, 4.5$ and $6$ are respectively in the small intervals $0.233<T_{\text{tr}}<0.250$, $0.323<T_{\text{tr}}<0.357$, $0.250<T_{\text{tr}}<0.270$ and $0.233<T_{\text{tr}}<0.250$. In the phase diagram (Fig.~\ref{fig:main:phasediagram}), the black and the blue lines represent the upper and lower boundaries of the transition temperatures determined by our DQMC simulations illustrated in Fig.~\ref{fig:main:D}. With denser data points, the two boundaries should merge into one. At $T = 0.294$, as the Hubbard $|U|$ increases, the system first undergoes a transition from the semimetal phase to the CDW phase and then enters the trion liquid phase caused by thermal melting of CDW order at strong coupling. As shown in Fig.~\ref{fig:main:phasediagram}, the transition temperature $T_{\text{tr}}(|U|)$ presents non-monotonic dependence on Hubbard attraction $|U|$, which can be understood by the second-order perturbation theory as follows.

\begin{figure}[t]
  \centering
  \includegraphics[width=0.9\linewidth]{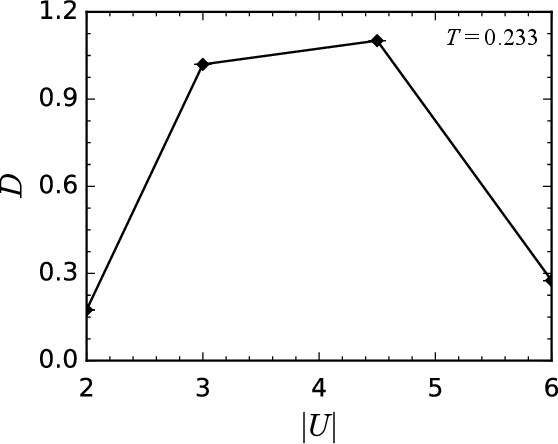}
  \caption{The CDW order parameter $D$ as a function of Hubbard $|U|$ at $T=0.233$. Error bars are smaller than the data points.
  } \label{fig:D-U}
\end{figure}

In the attractive SU(3) Hubbard model (Eq.~\eqref{main.Eq.1}), when $|U|$ is sufficiently large, one can treat the Hubbard interaction term as the unperturbed Hamiltonian and the hopping term as the perturbation. The second-order effective Hamiltonian can be derived as
\begin{equation}\label{eq:A3}
  H_{\text{rep}}=\frac{t^{2}}{2\left|U\right|}\sum_{\langle ij \rangle\alpha}n_{i\alpha}n_{j\alpha}
\end{equation}
which describes the effective repulsion between nearest-neighbor on-site trions \cite{Titvinidze2011}. At low temperatures, on-site trions tend to occupy the same sublattice minimizing the free energy, which develops CDW order. When $T>T_{\text{tr}}(|U|)$, the entropy contribution wins over the energy contribution, and to achieve minimum free energy, on-site trions tend to distribute randomly on the bipartite lattice, which gives rise to the thermal melting of the CDW order. The increase of $|U|$ decreases the energy scale $\frac{t^{2}}{|U|}$ and thus the CDW order can be thermally melted at even lower trsnsition temperatures. Hence, the transition temperature $T_{\text{tr}}(|U|)$ decreases with increasing $|U|$ in the strong-coupling regime, from which we can infer that the thermal CDW state can always be thermally melted into delocalized on-site trions at sufficiently strong coupling. This also implies the non-monotonic variation of the CDW order strength with Hubbard $|U|$ at constant temperature $T<\text{max}(T_{\text{tr}}(|U|))$. At $T=0.233$, the finite-size extrapolation of the CDW order strength for various Hubbard $|U|$ are presented in Fig.~\ref{fig:D-U}. The thermal fluctuation is the cause of the nonmonotonic behavior of the CDW ordering with increasing $|U|$, while at $T=0$ the CDW order strength increases monotonically with $|U|$.

\section{Trion formation}\label{sec:occupancy}

On-site trions are basically classical states since they constitute the ground state of the interaction term in Eq.~\eqref{main.Eq.1}, while off-site trions arising from quantum fluctuations are a direct consequence of the non-commutation between the hopping term and the interaction term in Eq.~\eqref{main.Eq.1}.
In the SU($2$) case, quantum fluctuations can be reflected by the temperature dependence of double occupancy \cite{Gorelik2010,Tang2013}, and in experiments the site-resolved imaging technique can be used to detect the on-site particle number occupation \cite{parsons2016site,cheuk2016observation,boll2016spin}.
In this section, we shall demonstrate the effects of on-site and off-site trions by simulating the temperature dependence of triple occupancy.
In QMC simulations, the triple occupancy is defined as
\begin{equation}
P_{3}(T,U)=\frac{1}{2L^{2}}\sum_{i} \langle n_{i1}n_{i2}n_{i3}\rangle.
\end{equation}
In the non-interacting/high-temperature limit, the three-color correlations can be decoupled and thus $\lim_{T\to\infty}P_{3}(T,U)=\lim_{U\to0}P_{3}(T,U)=0.125$.

\begin{figure}[t]
  \centering
  \includegraphics[width=0.96\linewidth]{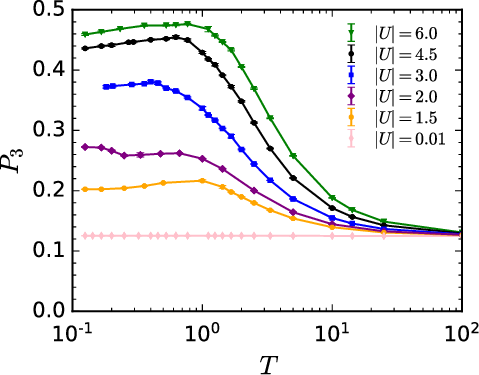}
  \caption{Triple occupancy $P_{3}$ as a function of temperature $T$ for various Hubbard $\left|U\right|$. The lattice size is $L=9$.
  } \label{fig:main:P3}
\end{figure}

The temperature dependence of triple occupancy $P_{3}$ for various Hubbard $|U|$ are plotted in Fig.~\ref{fig:main:P3}.
At very small $|U|=0.01$, $P_{3}$ is almost equal to the constant $0.125$, as expected. When $|U|=1.5$, unbound fermions dominate, but they tend to form trions for low temperatures $T < t$ so that formation of trions results in a reduction of the free energy. For $T > t$ trions break up into unbound fermions since maximizing entropy can lead to a minimum free energy. Thus the nonmonotonic temperature dependence of $P_{3}$ for $|U|=1.5$ reflects the roles for the energy and the entropy to minimize the free energy as the temperature varies. When $|U|=2$, the minimum of $P_{3}$ at $T = 0.25$ indicates the point where the trions start to replace the dominant role of unbound fermions as the temperature varies.
When $|U|=3, 4.5$ and $6$, each triple occupancy $P_{3}$ becomes a non-monotonic function of $T$ and achieves a peak at the temperatures $T^*$ which is slightly higher than the transition temperature $T_{\text{tr}}$. When $T > T^*$, $P_{3}$ decreases with increasing temperature and converges at the high-temperature limit $0.125$, regardless of the values of Hubbard $|U|$.

\begin{figure}[t]
  \centering
  \includegraphics[width=0.96\linewidth]{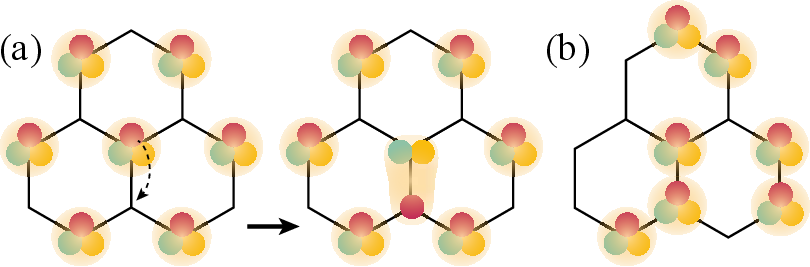}
  \caption{(a) The hopping process in the trion CDW state. (b) When on-site trions distribute randomly, a considerable percentage of such hopping processes are forbbiden by the Pauli exclusion.
  } \label{fig:main:trion01}
\end{figure}

The non-monotonic temperature dependence of $P_{3}$ for $|U|=3, 4.5$ and $6$ can be explained as follows. When the system is in the trion CDW states, as illustrated in Fig.~\ref{fig:main:trion01} (a), one fermion in each on-site trion can hop to three nearest-neighbor sites due to quantum fluctuations, which transforms an on-site trion into an off-site trion. This process reduces the triple occupancy $P_{3}$ to below $0.5$. At $T = T^{*}$, the CDW order has been thermally melted and on-site trions distribute randomly on the lattice. In this case, the nearest-neighbor sites of an on-site trion may be occupied by other on-site trions. Hence, the Pauli exclusion principle may freeze some channels of the nearest-neighbor hopping process, as shown in Fig.~\ref{fig:main:trion01}(b). As a result, the formation of off-site trions is suppressed by the random distribution of on-site trions, leading to the maximum of triple occupancy $P_{3}(T^{*})>P_{3}(T\rightarrow 0)$. With further increase of the temperature, trions start to break up into unbound fermions, so that $P_{3}$ decreases  and reaches to $0.125$ in the high-temperature limit.

\begin{figure}[t]
  \centering
  \includegraphics[width=0.96\linewidth]{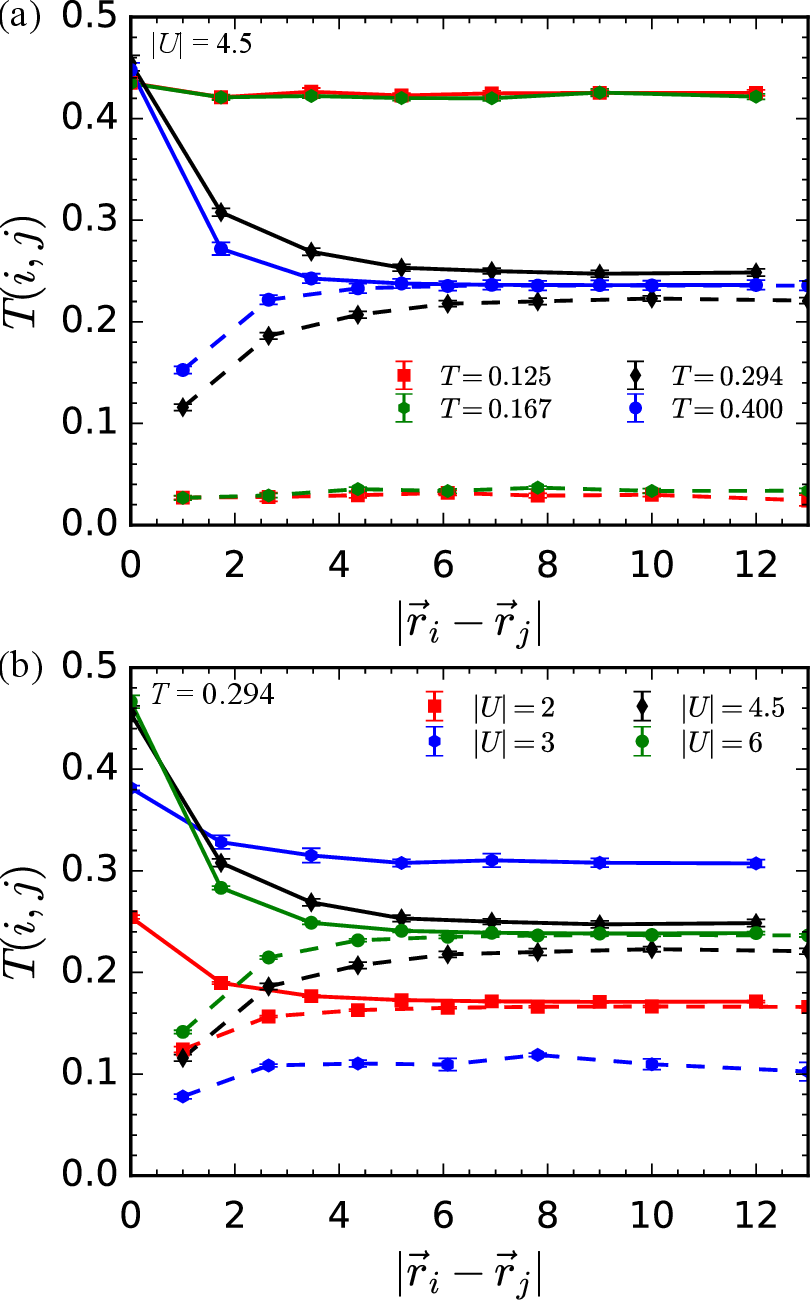}
  \caption{$T(i,j)$ as a function of $|\vec{r}_{i}-\vec{r}_{j}|$. (a) Varying temperature $T$ at constant Hubbard $\left| U\right|=4.5$.
  (b) Varying Hubbard $\left| U\right|$ at constant temperature $T=0.294$.  The solid and dashed curves represent respectively $T^{A}(i,j)$ and $T^{AB}(i,j)$. The lattice size is $L=9$.
  } \label{fig:main:Tij}
\end{figure}

The finite-temperature CDW transition is accompanied with the formation of trions and breaking of lattice inversion symmetry, which can also be manifested by the correlation function \cite{Kantian2009,Molina2009,Pohlmann2013,xu2019quantum},
\begin{equation}
  T(i,j)=\langle n_{i1}n_{i2}n_{j3}\rangle.
\end{equation}
It measures the correlation between the color-3 fermion at site $j$ and the color-1,2 fermions at site $i$. When $r_{ij}=|\vec{r}_{i}-\vec{r}_{j}|$ is sufficiently large, $T(i,j)$ depends on which sublattice the lattice sites $i$ and $j$ belong to, irrelevant to the length of $r_{ij}$, owing to the density uniformity within a sublattice. For convenience, $T(i,j)$ with $i,j$ being on the same sublattice and on different sublattices are respectively denoted by $T^{A}(i,j)$ and $T^{AB}(i,j)$.

In Fig.~\ref{fig:main:Tij} (a), the variation of $T(i,j)$ with $r_{ij}$ for various temperatures illustrates the thermal CDW phase transition at $|U|=4.5$.  For $T < 0.294$, $T^{A}(i,j)$ is much larger than $T^{AB}(i,j)$, which manifests the lattice inversion symmetry breaking of the solid-like CDW phase.
At $T=0.4$, $T^{A}(i,j)$ and $T^{AB}(i,j)$ converge to a common value 0.236, reflecting the lattice inversion symmetry of the trion liquid phase. Note that the converged value 0.236 is slightly smaller than $0.25$, which implies coexistence of majority on-site trions and minority off-site trions. The minimum of $T(i,j)$ appears at $r_{ij}=1$, which follows the behavior of $T(i,j)$ in the the large-$|U|$ limit due to the effective nearest-neighbor repulsion between on-site trions described by Eq.~\eqref{eq:A3}.

In Fig.~\ref{fig:main:Tij} (b), the interaction-induced phase transition is illustrated at $T=0.294$.
With increasing $|U|$, $T^{A}(i,j)$ and $T^{AB}(i,j)$ first converge to $0.17$ at $|U|=2$, and then have different values for $2<|U|<6$, and eventually converge to $0.24$ at $|U|=6$. This suggests that with increasing $|U|$, the system undergoes phase transitions from the semimetal phase to the CDW phase and thence to the trion liquid phase, which is consistent with the phase diagram (Fig.~\ref{fig:main:phasediagram}). At $|U|=2$, the converged value 0.17 is larger than 0.125 and much smaller than $0.25$, which manifests that a small number of unbound fermions form trions and thus the liquid-like phase consists of randomly distributed minority trions and majority unbound fermions. At $|U|=6$, the converged value 0.24 is slightly lower than $0.25$, implying that the liquid-like phase consists of randomly distributed majority on-site trions and minority off-site trions.

\section{The entropy-temperature relation}\label{sec:entropy}

\begin{figure}[b]
  \centering
  \includegraphics[width=0.96\linewidth]{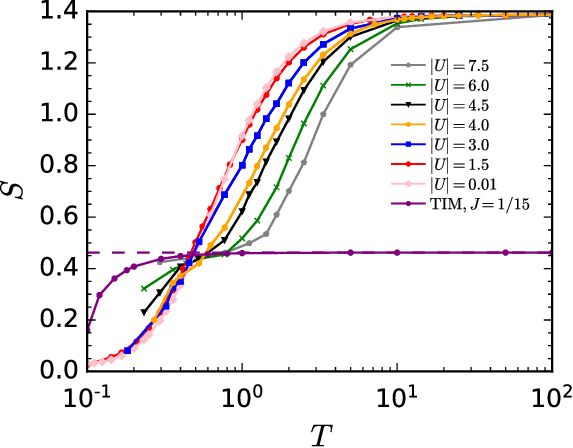}
  \caption{The entropy per particle $S$ as a function of temperature $T$ for various Hubbard $\left |U \right|$ (Lattice size $L=9$). The entropy-temperature relation of the trionic Ising model (TIM) is calculated by classical Monte Carlo simulations.
  } \label{fig:main:ST}
\end{figure}

We shall demonstrate how on-site trions and off-site trions affect the specific entropy, which is directly measurable in ultracold atom experiments \cite{Bloch2008}.
In QMC simulations, the entropy per particle can be calculated by using the following formula \cite{Zhou2017}:
\begin{equation} \label{eq:entropy}
  \frac{S(T)}{k_{B}}=\frac{S(\infty)}{k_{B}}+\frac{E(T)}{T}-\int^{\infty}_{T} dT'\frac{E(T')}{T'^{2}},
\end{equation}
where $E(T)$ is the total energy per particle and $S(\infty)$ is the entropy per particle in the high-temperature limit.
In our SU(3) model, there are eight possible states on each site in the high-temperature limit and thus $\frac{S(\infty)}{k_{\text{B}}}=\frac{\ln{8}}{1.5}=2\ln2$. In Eq.~\eqref{eq:entropy}, when $T'<1$, the numerical errors of $E(T')$ is amplified $\frac{1}{T'^{2}}$ times in the integral. Alternatively one can change the variable and obtain
\begin{equation} \label{eq:equa.1}
    \frac{S(T)}{k_{\text{B}}}=\frac{S(\infty)}{k_{\text{B}}}+\beta E(\beta)-\int_{0}^{\frac{1}{T}} d\beta'\ E(\beta')
\end{equation}
with inverse temperature $\beta=\frac{1}{T}$. When $T<1$, the numerical error in Eq.~(\ref{eq:equa.1}) is amplified $\beta$ times which is smaller than that in Eq.~(\ref{eq:entropy}).
Nevertheless, when $|U|>4$, the calculated values of $S(T)$ are reliable only for $\beta \leqslant 4.3$, as discussed in Appendix~\ref{app.error}.
Therefore, when plotting the temperature variation of specific entropy, specific entropies are only calculated for $T\gtrsim0.23$ when $|U|>4$.

Figure~\ref{fig:main:ST} shows that the small curve-crossing area characterizes the separation between low-temperature and high-temperature regions, and at low temperatures specific entropy $S$ increases monotonically with Hubbard $|U|$. We now analyze the $|U|$ dependence of the specific entropy at low temperatures and the  resulting consequence. At strong coupling where the system is in the trion CDW phase, the entropy $S$ is mainly contributed by the degrees of freedom of on-site trions.
With increasing $|U|$, since the effective nearest-neighbor repulsion between on-site trions decreases, on-site trions tend to be highly delocalized and then the system becomes less ordered (entropy increase). At weak coupling, the system is in the semimetal state, the entropy of which is mainly contributed by fermions near Dirac points and thus is small due to the vanishing density of states at half filling. Consequently the CDW state is less ordered than the thermal Dirac semimetal. The specific entropy of the system therefore increases with $|U|$ at low temperatures. As a consequence, the SU(3) fermions can be driven to lower temperatures by adiabatically increasing the strength of the attractive Hubbard interaction, exhibiting the Pomeranchuk effect. In the phase diagram (Fig.~\ref{fig:main:phasediagram}), the isoentropy curve of $S/k_B=0.31$ intersects the phase boundary, manifesting a possible scenario for the experimental realization of CDW states of attractive
SU(3) Dirac fermions via Pomeranchuk cooling.

\begin{figure}[t]
    \centering
    \includegraphics[width=0.96\linewidth]{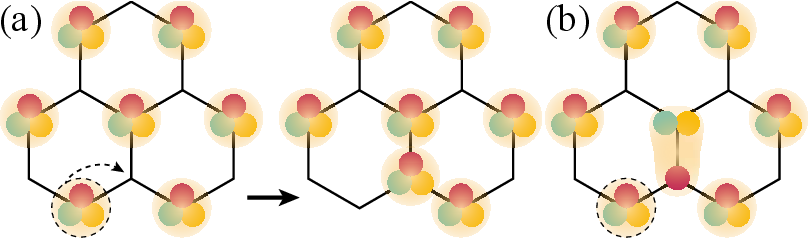}
    \caption{(a) The hopping process of on-site trions in the trionic Ising model. (b) In the attractive SU(3) Hubbard model, the off-site trion blocks some hopping channels of on-site trions due to the Pauli exclusion.
    } \label{fig:main:trion02}
  \end{figure}

The strong-coupling limit of our model is the trionic Ising model \cite{Titvinidze2011}, in which spin up (down) corresponds to an on-site trion (trionic hole) and the Ising coupling $J=\frac {t^{2}} {2|U|}$.
In Fig.~\ref{fig:main:ST}, the specific entropy $S$ of the trionic Ising model in the low-temperature region is larger than that of our model.
In the trionic Ising model, $S$ is contributed by the degrees of freedom of on-site trions.
However, in our model, due to the Pauli exclusion off-site trions block some hopping channels of on-site trions, which reduces entropy contributed by on-site trions, as illustrated in Fig.~\ref{fig:main:trion02}. This explains that the attractive SU(3) Hubbard model is more ordered than the trionic Ising model at low temperatures.

At high temperatures, trions break up into unbound fermions and thus the specific entropy is mainly contributed by the degrees of freedom of unbound fermions, which contributes more possible states than on-site trions.
Increasing $|U|$ favors the formation of on-site trions, and thus reduces the degrees of freedom of unbound fermions.
As a result, the specific entropy decreases with increasing $|U|$ at high temperatures.

When $|U|\geqslant 3$, a plateau-like regime appears in each $S-T$ curve and grows with increasing $|U|$, as shown in Fig.~\ref{fig:main:ST}. In fact, the roles of trions and unbound fermions in entropy production are separated at around the plateau-like regime. For larger $|U|$, the melted trions need to be heated up to an even higher temperature along the plateau until they break up into unbound fermions, which expands the plateau-like regime. For sufficiently large $|U|$ (i.e. trionic Ising model), a plateau of $S=(2\ln2)/3$ develops with increasing temperature, due to the full release of trion entropy.

\section{The density compressibility}\label{sec:knn}

\begin{figure}[b]
  \centering
  \includegraphics[width=0.96\linewidth]{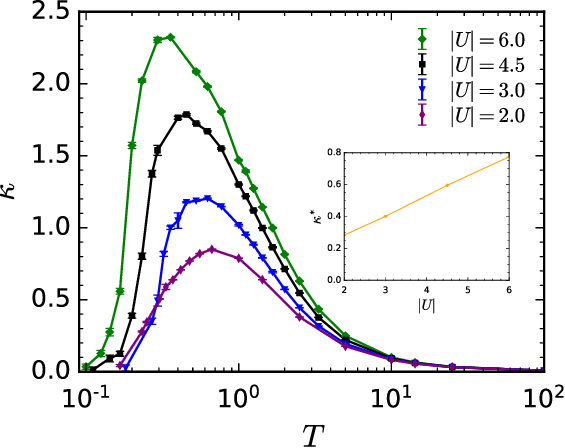}
  \caption{The density compressibility $\kappa$ as a function of temperature $T$ for various Hubbard $\left|U\right|$. The inset plots the peak value $\kappa^{*}(|U|)$ as a function of $|U|$. The lattice size is $L=9$.
  \label{fig:main:kappa}
  }
\end{figure}

The density compressibility is a physical observable in cold atom experiments as well. It is related to the global density fluctuations.
In our simulations, the density compressibility is defined as
\begin{equation}
\kappa=\frac{\beta}{2L^{2}}\left[\Big<\Big(\sum_{i}n_{i}\Big)^{2}\Big>-\Big<\sum_{i}n_{i}\Big>^{2}\right],
\end{equation}
where $i$ runs over the entire lattice.

Figure~\ref{fig:main:kappa} presents the density compressibility $\kappa$ as a function of temperature $T$ for various values of Hubbard $|U|\geqslant 2$.
At temperatures lower than the temperature of the curve-crossing point of $|U|=2$ and $|U|=3$, the $|U|$ dependence of $\kappa$ is obviously non-monotonic. When $2\leqslant|U|<3$, $\kappa$ decreases with increasing $|U|$, which manifests that unbound fermions are driven to form localized trions causing less compressible (harder) CDW state. When $3\leqslant|U|\leqslant6$, $\kappa$ increases with $|U|$. This implies that, as discussed in Sec.~\ref{sec:cdw-trans} and Sec.~\ref{sec:entropy}, trions tend to be more delocalized and thus charge fluctuations increase, leading to more compressible (softer) CDW state. Note that $\kappa$ for $|U|>3$ is significantly larger than that for $|U|<3$, because trions carry triple charge of an unbound fermion and enhance density fluctuations.
At constant Hubbard $|U|\geqslant 2$, the temperature dependence of $\kappa$ is nonmonotonic: at sufficiently low temperatures, $\kappa$ vanishes due to the insulating nature of the CDW phase, while at very high temperatures $\kappa$ behaves like that of a classical ideal gas, i.e., $\kappa(T)\sim\frac{1}{T}$ because trions break up into unbound fermions.
It is noteworthy that, when $|U|\geqslant 3$ (trion fluctuations dominate), the peaks of the $\kappa$-$T$ curves characterizes the melting temperature of CDW states where trion fluctuations are most significant and thus $\kappa$ reaches peak value. For sufficiently large $|U|$, as analyzed in Sec.~\ref{sec:cdw-trans}, the melting temperature is proportional to $\frac{1}{|U|}$ and consequently increasing $|U|$ moves the peak position towards lower temperature and also elevates the peak value (since $\kappa$ increases with $|U|$), leading to the $1/|U|$ divergence of $\kappa$ in the vicinity of zero temperature. This large-$U$ behavior even holds for $|U|\geqslant 3$, as shown in Fig.~\ref{fig:main:kappa}. In addition, the peak value of the density compressibility $\kappa^{*}(|U|)$ is nearly linear in $|U|$.

\section{Conclusions} \label{sec:conclusion}
We have performed DQMC simulations of the thermodynamic properties of the half-filled attractive SU(3) Hubbard model on a honeycomb lattice. We obtain the finite-temperature phase diagram in which the disordered phase and the CDW phase are separated by the phase boundary.
We have also investigated the influences of trions on the thermodynamic properties of the attractive SU(3) Dirac fermions by simulating the temperature dependence of triple occupancy, entropy and density compressibility.

When $|U|>|U_c|$ ($U_c = -1.52$ is the quantum critical point), lowering temperature can induce thermal CDW transitions. At constant temperature $T<\text{max}(T_{\text{tr}}(|U|))$, increasing Hubbard $|U|$ induces a semimetal-to-CDW transition at weak coupling where the density fluctuations of unbound fermions dominate, and further causes trion CDW state, and ultimately leads to a transition from the CDW to the trion liquid phase at sufficiently strong coupling where the density fluctuations of on-site trions govern. In the trion CDW region, isothermally increasing Hubbard $|U|$ decreases effective repulsion between on-site trions and thus enhances the delocalization of on-site trions, which leads to the increase in entropy (i.e. the Pomeranchuk effect) and the melting of the CDW state.

In the trion CDW region, off-site trions arise from quantum fluctuations - one fermion from an on-site trion hops to the nearest-neighbor site, forming an off-site trion. The formation of off-site trions and the delocalization of on-site trions develop in an opposite way due to the Pauli exclusion principle. Increasing temperature enhances the delocalization of on-site trions and thus suppress the formation of off-site trions, leading to the non-monotonic temperature dependence of triple occupancy.

\acknowledgments
This work is financially supported by the National
Natural Science Foundation of China under Grants No. 11874292, No. 11729402, and No. 11574238.
We acknowledge the support of the Supercomputing Center of Wuhan University.

\appendix

\section{The Hubbard-Stratonovich decomposition}
\label{decomposition}

In Ref.~\cite{xu2019quantum},  the Hubbard-Stratonovich decomposition of the interaction term in Eq.~(\ref{main.Eq.1}) is written as

\begin{equation}
\begin{aligned}
&e^{-\Delta_{\tau}\sum_{\alpha<\beta}U_{\alpha\beta}(n_{\alpha}-\frac{1}{2})(n_{\beta}-\frac{1}{2})}\\
&=\prod_{\alpha<\beta}e^{-\Delta_{\tau}U_{\alpha\beta}(n_{\alpha}-\frac{1}{2})(n_{\beta}-\frac{1}{2})}\\
&\approx\frac{1}{8}e^{-\frac{\Delta_{\tau}}{4}(U_{12}+U_{13}+U_{23})}\sum_{s_{\alpha\beta}=\pm 1}e^{\sum_{\alpha<\beta}s_{\alpha\beta}\lambda_{\alpha\beta}(c^{\dagger}_{\alpha}c_{\beta}-c_{\beta}^{\dagger}c_{\alpha})},
\end{aligned}
\end{equation}
with $\lambda_{\alpha\beta}=\arccos e^{\frac{\Delta_{\tau}U_{\alpha\beta}}{2}}$.
This decomposition is not exact since $c^{\dagger}_{1}c_{2}-c^{\dagger}_{2}c_{1}$, $c^{\dagger}_{1}c_{3}-c^{\dagger}_{3}c_{1}$ and $c^{\dagger}_{2}c_{3}-c^{\dagger}_{3}c_{2}$ do not commute with each other. In our QMC simulations, we adopt the exact decomposition below
\begin{equation}
\begin{aligned}
&e^{-\Delta_{\tau}\sum_{\alpha<\beta}U_{\alpha\beta}(n_{\alpha}-\frac{1}{2})(n_{\beta}-\frac{1}{2})}\\
&=\frac{1}{8}e^{-\frac{\Delta_{\tau}}{4}(U_{12}+U_{13}+U_{23})}\sum_{s_{\alpha\beta}=\pm 1}\prod_{\alpha<\beta}e^{s_{\alpha\beta}\lambda_{\alpha\beta}(c^{\dagger}_{\alpha}c_{\beta}-c_{\beta}^{\dagger}c_{\alpha})}.
\end{aligned}
\end{equation}

\section{The numerical error of the total energy per particle at low temperatures} \label{app.error}

\begin{figure}[tbh]
  \centering
  \includegraphics[width=0.96\linewidth]{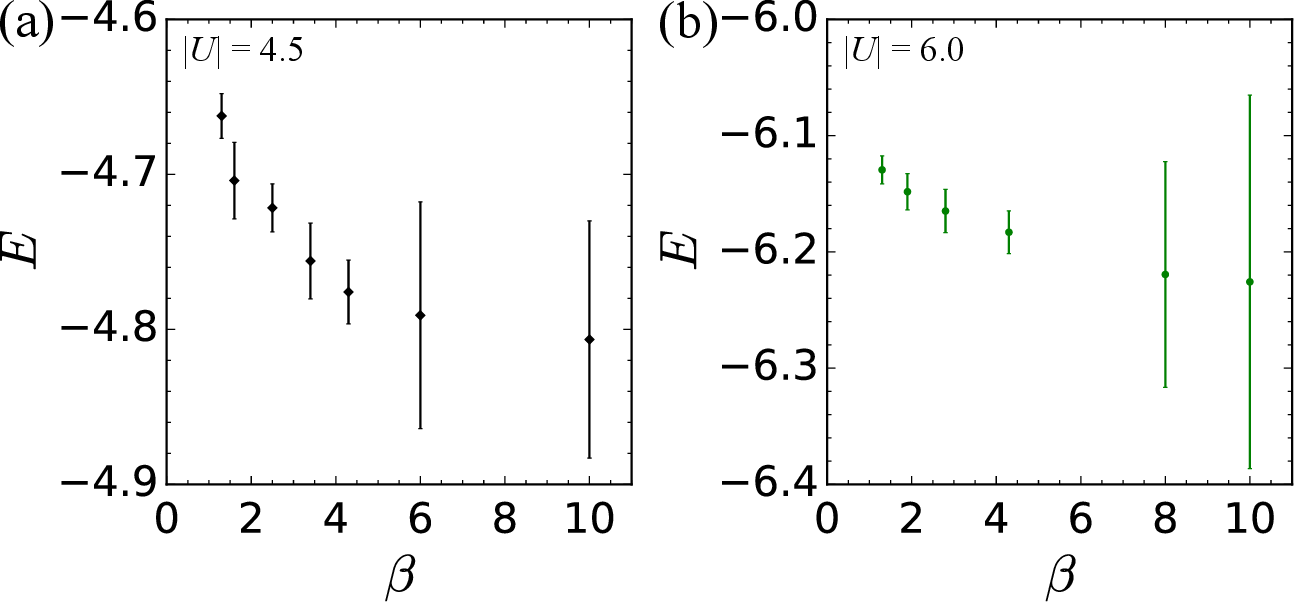}
  \caption{The total energy per particle $E$ versus inverse temperature $\beta$ at (a) $\left|U\right|=4.5$ and (b) $\left|U\right|=6.0$ respectively. The numerical errors $\Delta E(\beta)$ are replaced by $\beta \Delta E(\beta)$. The length of the line segments represents the value of $\beta \Delta E(\beta)$.
  } \label{fig:main:energy}
\end{figure}

According to Eq.~\eqref{eq:equa.1}, large inverse temperatures $\beta$ amplify the numerical error of the total energy per particle $\Delta E$.
Thus, $\Delta E$ can not be ignored when we calculate $S$ at low temperatures, especially for $|U|>4$.
In Figs.~\ref{fig:main:energy} (a) and (b), the numerical error is expressed by $\beta \Delta E(\beta)$ instead of $\Delta E(\beta)$ and the relations of the total energy per particle $E$ versus $\beta$ are presented for $|U|=4.5$ and $6.0$, respectively. At large $\beta$, the amplified numerical errors $\beta\Delta E$ are so large that the calculated values of $S$ become unreliable.

\begin{thebibliography}{28}%
\makeatletter
\providecommand \@ifxundefined [1]{%
 \@ifx{#1\undefined}
}%
\providecommand \@ifnum [1]{%
 \ifnum #1\expandafter \@firstoftwo
 \else \expandafter \@secondoftwo
 \fi
}%
\providecommand \@ifx [1]{%
 \ifx #1\expandafter \@firstoftwo
 \else \expandafter \@secondoftwo
 \fi
}%
\providecommand \natexlab [1]{#1}%
\providecommand \enquote  [1]{``#1''}%
\providecommand \bibnamefont  [1]{#1}%
\providecommand \bibfnamefont [1]{#1}%
\providecommand \citenamefont [1]{#1}%
\providecommand \href@noop [0]{\@secondoftwo}%
\providecommand \href [0]{\begingroup \@sanitize@url \@href}%
\providecommand \@href[1]{\@@startlink{#1}\@@href}%
\providecommand \@@href[1]{\endgroup#1\@@endlink}%
\providecommand \@sanitize@url [0]{\catcode `\\12\catcode `\$12\catcode
  `\&12\catcode `\#12\catcode `\^12\catcode `\_12\catcode `\%12\relax}%
\providecommand \@@startlink[1]{}%
\providecommand \@@endlink[0]{}%
\providecommand \url  [0]{\begingroup\@sanitize@url \@url }%
\providecommand \@url [1]{\endgroup\@href {#1}{\urlprefix }}%
\providecommand \urlprefix  [0]{URL }%
\providecommand \Eprint [0]{\href }%
\providecommand \doibase [0]{http://dx.doi.org/}%
\providecommand \selectlanguage [0]{\@gobble}%
\providecommand \bibinfo  [0]{\@secondoftwo}%
\providecommand \bibfield  [0]{\@secondoftwo}%
\providecommand \translation [1]{[#1]}%
\providecommand \BibitemOpen [0]{}%
\providecommand \bibitemStop [0]{}%
\providecommand \bibitemNoStop [0]{.\EOS\space}%
\providecommand \EOS [0]{\spacefactor3000\relax}%
\providecommand \BibitemShut  [1]{\csname bibitem#1\endcsname}%
\let\auto@bib@innerbib\@empty
\bibitem [{\citenamefont {Gorshkov}\ \emph {et~al.}(2010)\citenamefont
  {Gorshkov}, \citenamefont {Hermele}, \citenamefont {Gurarie}, \citenamefont
  {Xu}, \citenamefont {Julienne}, \citenamefont {Ye}, \citenamefont {Zoller},
  \citenamefont {Demler}, \citenamefont {Lukin},\ and\ \citenamefont
  {Rey}}]{Gorshkov2010}%
  \BibitemOpen
  \bibfield  {author} {\bibinfo {author} {\bibfnamefont {A.}~\bibnamefont
  {Gorshkov}}, \bibinfo {author} {\bibfnamefont {M.}~\bibnamefont {Hermele}},
  \bibinfo {author} {\bibfnamefont {V.}~\bibnamefont {Gurarie}}, \bibinfo
  {author} {\bibfnamefont {C.}~\bibnamefont {Xu}}, \bibinfo {author}
  {\bibfnamefont {P.}~\bibnamefont {Julienne}}, \bibinfo {author}
  {\bibfnamefont {J.}~\bibnamefont {Ye}}, \bibinfo {author} {\bibfnamefont
  {P.}~\bibnamefont {Zoller}}, \bibinfo {author} {\bibfnamefont
  {E.}~\bibnamefont {Demler}}, \bibinfo {author} {\bibfnamefont
  {M.}~\bibnamefont {Lukin}}, \ and\ \bibinfo {author} {\bibfnamefont
  {A.}~\bibnamefont {Rey}},\ }\href {https://www.nature.com/articles/nphys1535}
  {\bibfield  {journal} {\bibinfo  {journal} {Nat. Phys.}\ }\textbf {\bibinfo
  {volume} {6}},\ \bibinfo {pages} {289} (\bibinfo {year} {2010})}\BibitemShut
  {NoStop}%
\bibitem [{\citenamefont {Wu}(2012)}]{Wu2012}%
  \BibitemOpen
  \bibfield  {author} {\bibinfo {author} {\bibfnamefont {C.}~\bibnamefont
  {Wu}},\ }\href {https://www.nature.com/articles/nphys2432} {\bibfield
  {journal} {\bibinfo  {journal} {Nat. Phys.}\ }\textbf {\bibinfo {volume}
  {8}},\ \bibinfo {pages} {784} (\bibinfo {year} {2012})}\BibitemShut {NoStop}%
\bibitem [{\citenamefont {Taie}\ \emph {et~al.}(2010)\citenamefont {Taie},
  \citenamefont {Takasu}, \citenamefont {Sugawa}, \citenamefont {Yamazaki},
  \citenamefont {Tsujimoto}, \citenamefont {Murakami},\ and\ \citenamefont
  {Takahashi}}]{Taie2010}%
  \BibitemOpen
  \bibfield  {author} {\bibinfo {author} {\bibfnamefont {S.}~\bibnamefont
  {Taie}}, \bibinfo {author} {\bibfnamefont {Y.}~\bibnamefont {Takasu}},
  \bibinfo {author} {\bibfnamefont {S.}~\bibnamefont {Sugawa}}, \bibinfo
  {author} {\bibfnamefont {R.}~\bibnamefont {Yamazaki}}, \bibinfo {author}
  {\bibfnamefont {T.}~\bibnamefont {Tsujimoto}}, \bibinfo {author}
  {\bibfnamefont {R.}~\bibnamefont {Murakami}}, \ and\ \bibinfo {author}
  {\bibfnamefont {Y.}~\bibnamefont {Takahashi}},\ }\href {\doibase
  10.1103/PhysRevLett.105.190401} {\bibfield  {journal} {\bibinfo  {journal}
  {Phys. Rev. Lett.}\ }\textbf {\bibinfo {volume} {105}},\ \bibinfo {pages}
  {190401} (\bibinfo {year} {2010})}\BibitemShut {NoStop}%
\bibitem [{\citenamefont {DeSalvo}\ \emph {et~al.}(2010)\citenamefont
  {DeSalvo}, \citenamefont {Yan}, \citenamefont {Mickelson}, \citenamefont
  {Martinez~de Escobar},\ and\ \citenamefont {Killian}}]{Desalvo2010}%
  \BibitemOpen
  \bibfield  {author} {\bibinfo {author} {\bibfnamefont {B.~J.}\ \bibnamefont
  {DeSalvo}}, \bibinfo {author} {\bibfnamefont {M.}~\bibnamefont {Yan}},
  \bibinfo {author} {\bibfnamefont {P.~G.}\ \bibnamefont {Mickelson}}, \bibinfo
  {author} {\bibfnamefont {Y.~N.}\ \bibnamefont {Martinez~de Escobar}}, \ and\
  \bibinfo {author} {\bibfnamefont {T.~C.}\ \bibnamefont {Killian}},\ }\href
  {\doibase 10.1103/PhysRevLett.105.030402} {\bibfield  {journal} {\bibinfo
  {journal} {Phys. Rev. Lett.}\ }\textbf {\bibinfo {volume} {105}},\ \bibinfo
  {pages} {030402} (\bibinfo {year} {2010})}\BibitemShut {NoStop}%
\bibitem [{\citenamefont {Ottenstein}\ \emph {et~al.}(2008)\citenamefont
  {Ottenstein}, \citenamefont {Lompe}, \citenamefont {Kohnen}, \citenamefont
  {Wenz},\ and\ \citenamefont {Jochim}}]{Ottenstein2008}%
  \BibitemOpen
  \bibfield  {author} {\bibinfo {author} {\bibfnamefont {T.~B.}\ \bibnamefont
  {Ottenstein}}, \bibinfo {author} {\bibfnamefont {T.}~\bibnamefont {Lompe}},
  \bibinfo {author} {\bibfnamefont {M.}~\bibnamefont {Kohnen}}, \bibinfo
  {author} {\bibfnamefont {A.~N.}\ \bibnamefont {Wenz}}, \ and\ \bibinfo
  {author} {\bibfnamefont {S.}~\bibnamefont {Jochim}},\ }\href {\doibase
  10.1103/PhysRevLett.101.203202} {\bibfield  {journal} {\bibinfo  {journal}
  {Phys. Rev. Lett.}\ }\textbf {\bibinfo {volume} {101}},\ \bibinfo {pages}
  {203202} (\bibinfo {year} {2008})}\BibitemShut {NoStop}%
\bibitem [{\citenamefont {Huckans}\ \emph {et~al.}(2009)\citenamefont
  {Huckans}, \citenamefont {Williams}, \citenamefont {Hazlett}, \citenamefont
  {Stites},\ and\ \citenamefont {O'Hara}}]{Huckans2009}%
  \BibitemOpen
  \bibfield  {author} {\bibinfo {author} {\bibfnamefont {J.~H.}\ \bibnamefont
  {Huckans}}, \bibinfo {author} {\bibfnamefont {J.~R.}\ \bibnamefont
  {Williams}}, \bibinfo {author} {\bibfnamefont {E.~L.}\ \bibnamefont
  {Hazlett}}, \bibinfo {author} {\bibfnamefont {R.~W.}\ \bibnamefont {Stites}},
  \ and\ \bibinfo {author} {\bibfnamefont {K.~M.}\ \bibnamefont {O'Hara}},\
  }\href {\doibase 10.1103/PhysRevLett.102.165302} {\bibfield  {journal}
  {\bibinfo  {journal} {Phys. Rev. Lett.}\ }\textbf {\bibinfo {volume} {102}},\
  \bibinfo {pages} {165302} (\bibinfo {year} {2009})}\BibitemShut {NoStop}%
\bibitem [{\citenamefont {Wu}(2010)}]{Wu2010}%
  \BibitemOpen
  \bibfield  {author} {\bibinfo {author} {\bibfnamefont {C.}~\bibnamefont
  {Wu}},\ }\href {https://physics.aps.org/articles/v3/92} {\bibfield  {journal}
  {\bibinfo  {journal} {Physics}\ }\textbf {\bibinfo {volume} {3}},\ \bibinfo
  {pages} {92} (\bibinfo {year} {2010})}\BibitemShut {NoStop}%
\bibitem [{\citenamefont {Cai}\ \emph {et~al.}(2013)\citenamefont {Cai},
  \citenamefont {Hung}, \citenamefont {Wang}, \citenamefont {Zheng},\ and\
  \citenamefont {Wu}}]{Cai2013}%
  \BibitemOpen
  \bibfield  {author} {\bibinfo {author} {\bibfnamefont {Z.}~\bibnamefont
  {Cai}}, \bibinfo {author} {\bibfnamefont {H.-h.}\ \bibnamefont {Hung}},
  \bibinfo {author} {\bibfnamefont {L.}~\bibnamefont {Wang}}, \bibinfo {author}
  {\bibfnamefont {D.}~\bibnamefont {Zheng}}, \ and\ \bibinfo {author}
  {\bibfnamefont {C.}~\bibnamefont {Wu}},\ }\href {\doibase
  10.1103/PhysRevLett.110.220401} {\bibfield  {journal} {\bibinfo  {journal}
  {Phys. Rev. Lett.}\ }\textbf {\bibinfo {volume} {110}},\ \bibinfo {pages}
  {220401} (\bibinfo {year} {2013})}\BibitemShut {NoStop}%
\bibitem [{\citenamefont {Wang}\ \emph {et~al.}(2014)\citenamefont {Wang},
  \citenamefont {Li}, \citenamefont {Cai}, \citenamefont {Zhou}, \citenamefont
  {Wang},\ and\ \citenamefont {Wu}}]{Wang2014}%
  \BibitemOpen
  \bibfield  {author} {\bibinfo {author} {\bibfnamefont {D.}~\bibnamefont
  {Wang}}, \bibinfo {author} {\bibfnamefont {Y.}~\bibnamefont {Li}}, \bibinfo
  {author} {\bibfnamefont {Z.}~\bibnamefont {Cai}}, \bibinfo {author}
  {\bibfnamefont {Z.}~\bibnamefont {Zhou}}, \bibinfo {author} {\bibfnamefont
  {Y.}~\bibnamefont {Wang}}, \ and\ \bibinfo {author} {\bibfnamefont
  {C.}~\bibnamefont {Wu}},\ }\href {\doibase 10.1103/PhysRevLett.112.156403}
  {\bibfield  {journal} {\bibinfo  {journal} {Phys. Rev. Lett.}\ }\textbf
  {\bibinfo {volume} {112}},\ \bibinfo {pages} {156403} (\bibinfo {year}
  {2014})}\BibitemShut {NoStop}%
\bibitem [{\citenamefont {Zhou}\ \emph {et~al.}(2014)\citenamefont {Zhou},
  \citenamefont {Cai}, \citenamefont {Wu},\ and\ \citenamefont
  {Wang}}]{Zhou2014}%
  \BibitemOpen
  \bibfield  {author} {\bibinfo {author} {\bibfnamefont {Z.}~\bibnamefont
  {Zhou}}, \bibinfo {author} {\bibfnamefont {Z.}~\bibnamefont {Cai}}, \bibinfo
  {author} {\bibfnamefont {C.}~\bibnamefont {Wu}}, \ and\ \bibinfo {author}
  {\bibfnamefont {Y.}~\bibnamefont {Wang}},\ }\href {\doibase
  10.1103/PhysRevB.90.235139} {\bibfield  {journal} {\bibinfo  {journal} {Phys.
  Rev. B}\ }\textbf {\bibinfo {volume} {90}},\ \bibinfo {pages} {235139}
  (\bibinfo {year} {2014})}\BibitemShut {NoStop}%
\bibitem [{\citenamefont {Zhou}\ \emph {et~al.}(2016)\citenamefont {Zhou},
  \citenamefont {Wang}, \citenamefont {Meng}, \citenamefont {Wang},\ and\
  \citenamefont {Wu}}]{Zhou2016}%
  \BibitemOpen
  \bibfield  {author} {\bibinfo {author} {\bibfnamefont {Z.}~\bibnamefont
  {Zhou}}, \bibinfo {author} {\bibfnamefont {D.}~\bibnamefont {Wang}}, \bibinfo
  {author} {\bibfnamefont {Z.~Y.}\ \bibnamefont {Meng}}, \bibinfo {author}
  {\bibfnamefont {Y.}~\bibnamefont {Wang}}, \ and\ \bibinfo {author}
  {\bibfnamefont {C.}~\bibnamefont {Wu}},\ }\href {\doibase
  10.1103/PhysRevB.93.245157} {\bibfield  {journal} {\bibinfo  {journal} {Phys.
  Rev. B}\ }\textbf {\bibinfo {volume} {93}},\ \bibinfo {pages} {245157}
  (\bibinfo {year} {2016})}\BibitemShut {NoStop}%
\bibitem [{\citenamefont {Zhou}\ \emph {et~al.}(2017)\citenamefont {Zhou},
  \citenamefont {Wang}, \citenamefont {Wu},\ and\ \citenamefont
  {Wang}}]{Zhou2017}%
  \BibitemOpen
  \bibfield  {author} {\bibinfo {author} {\bibfnamefont {Z.}~\bibnamefont
  {Zhou}}, \bibinfo {author} {\bibfnamefont {D.}~\bibnamefont {Wang}}, \bibinfo
  {author} {\bibfnamefont {C.}~\bibnamefont {Wu}}, \ and\ \bibinfo {author}
  {\bibfnamefont {Y.}~\bibnamefont {Wang}},\ }\href {\doibase
  10.1103/PhysRevB.95.085128} {\bibfield  {journal} {\bibinfo  {journal} {Phys.
  Rev. B}\ }\textbf {\bibinfo {volume} {95}},\ \bibinfo {pages} {085128}
  (\bibinfo {year} {2017})}\BibitemShut {NoStop}%
\bibitem [{\citenamefont {Zhou}\ \emph {et~al.}(2018)\citenamefont {Zhou},
  \citenamefont {Wu},\ and\ \citenamefont {Wang}}]{Zhou2018}%
  \BibitemOpen
  \bibfield  {author} {\bibinfo {author} {\bibfnamefont {Z.}~\bibnamefont
  {Zhou}}, \bibinfo {author} {\bibfnamefont {C.}~\bibnamefont {Wu}}, \ and\
  \bibinfo {author} {\bibfnamefont {Y.}~\bibnamefont {Wang}},\ }\href {\doibase
  10.1103/PhysRevB.97.195122} {\bibfield  {journal} {\bibinfo  {journal} {Phys.
  Rev. B}\ }\textbf {\bibinfo {volume} {97}},\ \bibinfo {pages} {195122}
  (\bibinfo {year} {2018})}\BibitemShut {NoStop}%
\bibitem [{\citenamefont {Gorelik}\ and\ \citenamefont
  {Bl\"umer}(2009)}]{Gorelik2009}%
  \BibitemOpen
  \bibfield  {author} {\bibinfo {author} {\bibfnamefont {E.~V.}\ \bibnamefont
  {Gorelik}}\ and\ \bibinfo {author} {\bibfnamefont {N.}~\bibnamefont
  {Bl\"umer}},\ }\href {\doibase 10.1103/PhysRevA.80.051602} {\bibfield
  {journal} {\bibinfo  {journal} {Phys. Rev. A}\ }\textbf {\bibinfo {volume}
  {80}},\ \bibinfo {pages} {051602} (\bibinfo {year} {2009})}\BibitemShut
  {NoStop}%
\bibitem [{\citenamefont {Inaba}\ \emph {et~al.}(2010)\citenamefont {Inaba},
  \citenamefont {Miyatake},\ and\ \citenamefont {Suga}}]{Inaba2010}%
  \BibitemOpen
  \bibfield  {author} {\bibinfo {author} {\bibfnamefont {K.}~\bibnamefont
  {Inaba}}, \bibinfo {author} {\bibfnamefont {S.-y.}\ \bibnamefont {Miyatake}},
  \ and\ \bibinfo {author} {\bibfnamefont {S.-i.}\ \bibnamefont {Suga}},\
  }\href {\doibase 10.1103/PhysRevA.82.051602} {\bibfield  {journal} {\bibinfo
  {journal} {Phys. Rev. A}\ }\textbf {\bibinfo {volume} {82}},\ \bibinfo
  {pages} {051602} (\bibinfo {year} {2010})}\BibitemShut {NoStop}%
\bibitem [{\citenamefont {Miyatake}\ \emph {et~al.}(2010)\citenamefont
  {Miyatake}, \citenamefont {Inaba},\ and\ \citenamefont
  {Suga}}]{Miyatake2010}%
  \BibitemOpen
  \bibfield  {author} {\bibinfo {author} {\bibfnamefont {S.-y.}\ \bibnamefont
  {Miyatake}}, \bibinfo {author} {\bibfnamefont {K.}~\bibnamefont {Inaba}}, \
  and\ \bibinfo {author} {\bibfnamefont {S.-i.}\ \bibnamefont {Suga}},\ }\href
  {\doibase 10.1103/PhysRevA.81.021603} {\bibfield  {journal} {\bibinfo
  {journal} {Phys. Rev. A}\ }\textbf {\bibinfo {volume} {81}},\ \bibinfo
  {pages} {021603} (\bibinfo {year} {2010})}\BibitemShut {NoStop}%
\bibitem [{\citenamefont {Inaba}\ and\ \citenamefont {Suga}(2012)}]{Inaba2012}%
  \BibitemOpen
  \bibfield  {author} {\bibinfo {author} {\bibfnamefont {K.}~\bibnamefont
  {Inaba}}\ and\ \bibinfo {author} {\bibfnamefont {S.-i.}\ \bibnamefont
  {Suga}},\ }\href {\doibase 10.1103/PhysRevLett.108.255301} {\bibfield
  {journal} {\bibinfo  {journal} {Phys. Rev. Lett.}\ }\textbf {\bibinfo
  {volume} {108}},\ \bibinfo {pages} {255301} (\bibinfo {year}
  {2012})}\BibitemShut {NoStop}%
\bibitem [{\citenamefont {Inaba}\ and\ \citenamefont {Suga}(2013)}]{Inaba2013}%
  \BibitemOpen
  \bibfield  {author} {\bibinfo {author} {\bibfnamefont {K.}~\bibnamefont
  {Inaba}}\ and\ \bibinfo {author} {\bibfnamefont {S.}~\bibnamefont {Suga}},\
  }\href {https://www.worldscientific.com/doi/abs/10.1142/S0217984913300081}
  {\bibfield  {journal} {\bibinfo  {journal} {Mod. Phys. Lett. B}\ }\textbf
  {\bibinfo {volume} {27}},\ \bibinfo {pages} {1330008} (\bibinfo {year}
  {2013})}\BibitemShut {NoStop}%
\bibitem [{\citenamefont {Okanami}\ \emph {et~al.}(2014)\citenamefont
  {Okanami}, \citenamefont {Takemori},\ and\ \citenamefont
  {Koga}}]{Okanami2014}%
  \BibitemOpen
  \bibfield  {author} {\bibinfo {author} {\bibfnamefont {Y.}~\bibnamefont
  {Okanami}}, \bibinfo {author} {\bibfnamefont {N.}~\bibnamefont {Takemori}}, \
  and\ \bibinfo {author} {\bibfnamefont {A.}~\bibnamefont {Koga}},\ }\href
  {\doibase 10.1103/PhysRevA.89.053622} {\bibfield  {journal} {\bibinfo
  {journal} {Phys. Rev. A}\ }\textbf {\bibinfo {volume} {89}},\ \bibinfo
  {pages} {053622} (\bibinfo {year} {2014})}\BibitemShut {NoStop}%
\bibitem [{\citenamefont {Suga}(2015)}]{Suga2015}%
  \BibitemOpen
  \bibfield  {author} {\bibinfo {author} {\bibfnamefont {S.-i.}\ \bibnamefont
  {Suga}},\ }\href {\doibase 10.1103/PhysRevA.92.023617} {\bibfield  {journal}
  {\bibinfo  {journal} {Phys. Rev. A}\ }\textbf {\bibinfo {volume} {92}},\
  \bibinfo {pages} {023617} (\bibinfo {year} {2015})}\BibitemShut {NoStop}%
\bibitem [{\citenamefont {Yanatori}\ and\ \citenamefont
  {Koga}(2016{\natexlab{a}})}]{Yanatori2016}%
  \BibitemOpen
  \bibfield  {author} {\bibinfo {author} {\bibfnamefont {H.}~\bibnamefont
  {Yanatori}}\ and\ \bibinfo {author} {\bibfnamefont {A.}~\bibnamefont
  {Koga}},\ }\href {\doibase 10.1103/PhysRevB.94.041110} {\bibfield  {journal}
  {\bibinfo  {journal} {Phys. Rev. B}\ }\textbf {\bibinfo {volume} {94}},\
  \bibinfo {pages} {041110} (\bibinfo {year} {2016}{\natexlab{a}})}\BibitemShut
  {NoStop}%
\bibitem [{\citenamefont {Yanatori}\ and\ \citenamefont
  {Koga}(2016{\natexlab{b}})}]{Yanatori2016b}%
  \BibitemOpen
  \bibfield  {author} {\bibinfo {author} {\bibfnamefont {H.}~\bibnamefont
  {Yanatori}}\ and\ \bibinfo {author} {\bibfnamefont {A.}~\bibnamefont
  {Koga}},\ }\href {\doibase 10.7566/JPSJ.85.014002} {\bibfield  {journal}
  {\bibinfo  {journal} {J. Phys. Soc. Jpn.}\ }\textbf {\bibinfo {volume}
  {85}},\ \bibinfo {pages} {014002} (\bibinfo {year}
  {2016}{\natexlab{b}})}\BibitemShut {NoStop}%
\bibitem [{\citenamefont {Hasunuma}\ \emph {et~al.}(2016)\citenamefont
  {Hasunuma}, \citenamefont {Kaneko}, \citenamefont {Miyakoshi},\ and\
  \citenamefont {Ohta}}]{Hasunuma2016}%
  \BibitemOpen
  \bibfield  {author} {\bibinfo {author} {\bibfnamefont {T.}~\bibnamefont
  {Hasunuma}}, \bibinfo {author} {\bibfnamefont {T.}~\bibnamefont {Kaneko}},
  \bibinfo {author} {\bibfnamefont {S.}~\bibnamefont {Miyakoshi}}, \ and\
  \bibinfo {author} {\bibfnamefont {Y.}~\bibnamefont {Ohta}},\ }\href {\doibase
  10.7566/JPSJ.85.074704} {\bibfield  {journal} {\bibinfo  {journal} {J. Phys.
  Soc. Jpn.}\ }\textbf {\bibinfo {volume} {85}},\ \bibinfo {pages} {074704}
  (\bibinfo {year} {2016})}\BibitemShut {NoStop}%
\bibitem [{\citenamefont {Pohlmann}\ \emph {et~al.}(2013)\citenamefont
  {Pohlmann}, \citenamefont {Privitera}, \citenamefont {Titvinidze},\ and\
  \citenamefont {Hofstetter}}]{Pohlmann2013}%
  \BibitemOpen
  \bibfield  {author} {\bibinfo {author} {\bibfnamefont {J.}~\bibnamefont
  {Pohlmann}}, \bibinfo {author} {\bibfnamefont {A.}~\bibnamefont {Privitera}},
  \bibinfo {author} {\bibfnamefont {I.}~\bibnamefont {Titvinidze}}, \ and\
  \bibinfo {author} {\bibfnamefont {W.}~\bibnamefont {Hofstetter}},\ }\href
  {\doibase 10.1103/PhysRevA.87.023617} {\bibfield  {journal} {\bibinfo
  {journal} {Phys. Rev. A}\ }\textbf {\bibinfo {volume} {87}},\ \bibinfo
  {pages} {023617} (\bibinfo {year} {2013})}\BibitemShut {NoStop}%
\bibitem [{\citenamefont {Xu}\ \emph {et~al.}(2023)\citenamefont {Xu},
  \citenamefont {Li}, \citenamefont {Zhou}, \citenamefont {Wang}, \citenamefont
  {Wang}, \citenamefont {Wu},\ and\ \citenamefont
  {Wang}}]{xu2019quantum}%
  \BibitemOpen
  \bibfield  {author} {\bibinfo {author} {\bibfnamefont {H.}~\bibnamefont
  {Xu}}, \bibinfo {author} {\bibfnamefont {X.}~\bibnamefont {Li}}, \bibinfo
  {author} {\bibfnamefont {Z.}~\bibnamefont {Zhou}}, \bibinfo {author}
  {\bibfnamefont {X.}~\bibnamefont {Wang}}, \bibinfo {author} {\bibfnamefont
  {L.}~\bibnamefont {Wang}}, \bibinfo {author} {\bibfnamefont {C.}~\bibnamefont
  {Wu}}, \ and\ \bibinfo {author} {\bibfnamefont {Y.}~\bibnamefont {Wang}},\
  }\href {\doibase 10.1103/PhysRevResearch.5.023180} {\bibfield  {journal}
  {\bibinfo  {journal} {Phys. Rev. Res.}\ }\textbf {\bibinfo {volume} {5}},\
  \bibinfo {pages} {023180} (\bibinfo {year} {2023})}\BibitemShut {NoStop}%
\bibitem [{\citenamefont {Rapp}\ \emph {et~al.}(2007)\citenamefont {Rapp},
  \citenamefont {Zar\'and}, \citenamefont {Honerkamp},\ and\ \citenamefont
  {Hofstetter}}]{Rapp2007}%
  \BibitemOpen
  \bibfield  {author} {\bibinfo {author} {\bibfnamefont {A.}~\bibnamefont
  {Rapp}}, \bibinfo {author} {\bibfnamefont {G.}~\bibnamefont {Zar\'and}},
  \bibinfo {author} {\bibfnamefont {C.}~\bibnamefont {Honerkamp}}, \ and\
  \bibinfo {author} {\bibfnamefont {W.}~\bibnamefont {Hofstetter}},\ }\href
  {\doibase 10.1103/PhysRevLett.98.160405} {\bibfield  {journal} {\bibinfo
  {journal} {Phys. Rev. Lett.}\ }\textbf {\bibinfo {volume} {98}},\ \bibinfo
  {pages} {160405} (\bibinfo {year} {2007})}\BibitemShut {NoStop}%
\bibitem [{\citenamefont {Rapp}\ \emph {et~al.}(2008)\citenamefont {Rapp},
  \citenamefont {Hofstetter},\ and\ \citenamefont {Zar\'and}}]{Rapp2008}%
  \BibitemOpen
  \bibfield  {author} {\bibinfo {author} {\bibfnamefont {A.}~\bibnamefont
  {Rapp}}, \bibinfo {author} {\bibfnamefont {W.}~\bibnamefont {Hofstetter}}, \
  and\ \bibinfo {author} {\bibfnamefont {G.}~\bibnamefont {Zar\'and}},\ }\href
  {\doibase 10.1103/PhysRevB.77.144520} {\bibfield  {journal} {\bibinfo
  {journal} {Phys. Rev. B}\ }\textbf {\bibinfo {volume} {77}},\ \bibinfo
  {pages} {144520} (\bibinfo {year} {2008})}\BibitemShut {NoStop}%
\bibitem [{\citenamefont {Inaba}\ and\ \citenamefont {Suga}(2009)}]{Inaba2009}%
  \BibitemOpen
  \bibfield  {author} {\bibinfo {author} {\bibfnamefont {K.}~\bibnamefont
  {Inaba}}\ and\ \bibinfo {author} {\bibfnamefont {S.-i.}\ \bibnamefont
  {Suga}},\ }\href {\doibase 10.1103/PhysRevA.80.041602} {\bibfield  {journal}
  {\bibinfo  {journal} {Phys. Rev. A}\ }\textbf {\bibinfo {volume} {80}},\
  \bibinfo {pages} {041602} (\bibinfo {year} {2009})}\BibitemShut {NoStop}%
\bibitem [{\citenamefont {Inaba}\ and\ \citenamefont
  {Suga}(2011)}]{inaba2011color}%
  \BibitemOpen
  \bibfield  {author} {\bibinfo {author} {\bibfnamefont {K.}~\bibnamefont
  {Inaba}}\ and\ \bibinfo {author} {\bibfnamefont {S.-i.}\ \bibnamefont
  {Suga}},\ }\href {\doibase 10.1142/S021798491102670X} {\bibfield  {journal}
  {\bibinfo  {journal} {Mod. Phys. Lett. B}\ }\textbf {\bibinfo {volume}
  {25}},\ \bibinfo {pages} {987} (\bibinfo {year} {2011})}\BibitemShut
  {NoStop}%
\bibitem [{\citenamefont {Titvinidze}\ \emph {et~al.}(2011)\citenamefont
  {Titvinidze}, \citenamefont {Privitera}, \citenamefont {Chang}, \citenamefont
  {Diehl}, \citenamefont {Baranov}, \citenamefont {Daley},\ and\ \citenamefont
  {Hofstetter}}]{Titvinidze2011}%
  \BibitemOpen
  \bibfield  {author} {\bibinfo {author} {\bibfnamefont {I.}~\bibnamefont
  {Titvinidze}}, \bibinfo {author} {\bibfnamefont {A.}~\bibnamefont
  {Privitera}}, \bibinfo {author} {\bibfnamefont {S.-Y.}\ \bibnamefont
  {Chang}}, \bibinfo {author} {\bibfnamefont {S.}~\bibnamefont {Diehl}},
  \bibinfo {author} {\bibfnamefont {M.~A.}\ \bibnamefont {Baranov}}, \bibinfo
  {author} {\bibfnamefont {A.}~\bibnamefont {Daley}}, \ and\ \bibinfo {author}
  {\bibfnamefont {W.}~\bibnamefont {Hofstetter}},\ }\href
  {https://iopscience.iop.org/article/10.1088/1367-2630/13/3/035013} {\bibfield
   {journal} {\bibinfo  {journal} {New J. Phys.}\ }\textbf {\bibinfo {volume}
  {13}},\ \bibinfo {pages} {035013} (\bibinfo {year} {2011})}\BibitemShut
  {NoStop}%
\bibitem [{\citenamefont {Koga}\ and\ \citenamefont
  {Yanatori}(2017)}]{Koga2017}%
  \BibitemOpen
  \bibfield  {author} {\bibinfo {author} {\bibfnamefont {A.}~\bibnamefont
  {Koga}}\ and\ \bibinfo {author} {\bibfnamefont {H.}~\bibnamefont
  {Yanatori}},\ }\href {https://journals.jps.jp/doi/10.7566/JPSJ.86.034702}
  {\bibfield  {journal} {\bibinfo  {journal} {J. Phys. Soc. Jpn}\ }\textbf
  {\bibinfo {volume} {86}},\ \bibinfo {pages} {034702} (\bibinfo {year}
  {2017})}\BibitemShut {NoStop}%
\bibitem [{\citenamefont {Fodor}\ and\ \citenamefont {Katz}(2002)}]{Fodor2002}%
  \BibitemOpen
  \bibfield  {author} {\bibinfo {author} {\bibfnamefont {Z.}~\bibnamefont
  {Fodor}}\ and\ \bibinfo {author} {\bibfnamefont {S.~D.}\ \bibnamefont
  {Katz}},\ }\href {https://doi.org/10.1088%2F1126-6708%2F2002%2F03%2F014}
  {\bibfield  {journal} {\bibinfo  {journal} {J. High Energy Phys.}\ }\textbf
  {\bibinfo {volume} {2002}},\ \bibinfo {pages} {014} (\bibinfo {year}
  {2002})}\BibitemShut {NoStop}%
\bibitem [{\citenamefont {Aoki}\ \emph {et~al.}(2006)\citenamefont {Aoki},
  \citenamefont {Endr{\H{o}}di}, \citenamefont {Fodor}, \citenamefont {Katz},\
  and\ \citenamefont {Szabo}}]{Aoki2006}%
  \BibitemOpen
  \bibfield  {author} {\bibinfo {author} {\bibfnamefont {Y.}~\bibnamefont
  {Aoki}}, \bibinfo {author} {\bibfnamefont {G.}~\bibnamefont {Endr{\H{o}}di}},
  \bibinfo {author} {\bibfnamefont {Z.}~\bibnamefont {Fodor}}, \bibinfo
  {author} {\bibfnamefont {S.}~\bibnamefont {Katz}}, \ and\ \bibinfo {author}
  {\bibfnamefont {K.}~\bibnamefont {Szabo}},\ }\href {\doibase
  10.1038/nature05120} {\bibfield  {journal} {\bibinfo  {journal} {Nature}\
  }\textbf {\bibinfo {volume} {443}},\ \bibinfo {pages} {675} (\bibinfo {year}
  {2006})}\BibitemShut {NoStop}%
\bibitem [{\citenamefont {Wilczek}(2007)}]{Wilczek2007}%
  \BibitemOpen
  \bibfield  {author} {\bibinfo {author} {\bibfnamefont {F.}~\bibnamefont
  {Wilczek}},\ }\href {\doibase 10.1038/nphys635} {\bibfield  {journal}
  {\bibinfo  {journal} {Nat. Phys.}\ }\textbf {\bibinfo {volume} {3}},\
  \bibinfo {pages} {375} (\bibinfo {year} {2007})}\BibitemShut {NoStop}%
\bibitem [{\citenamefont {Kantian}\ \emph {et~al.}(2009)\citenamefont
  {Kantian}, \citenamefont {Dalmonte}, \citenamefont {Diehl}, \citenamefont
  {Hofstetter}, \citenamefont {Zoller},\ and\ \citenamefont
  {Daley}}]{Kantian2009}%
  \BibitemOpen
  \bibfield  {author} {\bibinfo {author} {\bibfnamefont {A.}~\bibnamefont
  {Kantian}}, \bibinfo {author} {\bibfnamefont {M.}~\bibnamefont {Dalmonte}},
  \bibinfo {author} {\bibfnamefont {S.}~\bibnamefont {Diehl}}, \bibinfo
  {author} {\bibfnamefont {W.}~\bibnamefont {Hofstetter}}, \bibinfo {author}
  {\bibfnamefont {P.}~\bibnamefont {Zoller}}, \ and\ \bibinfo {author}
  {\bibfnamefont {A.~J.}\ \bibnamefont {Daley}},\ }\href {\doibase
  10.1103/PhysRevLett.103.240401} {\bibfield  {journal} {\bibinfo  {journal}
  {Phys. Rev. Lett.}\ }\textbf {\bibinfo {volume} {103}},\ \bibinfo {pages}
  {240401} (\bibinfo {year} {2009})}\BibitemShut {NoStop}%
\bibitem [{\citenamefont {Azaria}\ \emph {et~al.}(2009)\citenamefont {Azaria},
  \citenamefont {Capponi},\ and\ \citenamefont {Lecheminant}}]{Capponi2009}%
  \BibitemOpen
  \bibfield  {author} {\bibinfo {author} {\bibfnamefont {P.}~\bibnamefont
  {Azaria}}, \bibinfo {author} {\bibfnamefont {S.}~\bibnamefont {Capponi}}, \
  and\ \bibinfo {author} {\bibfnamefont {P.}~\bibnamefont {Lecheminant}},\
  }\href {\doibase 10.1103/PhysRevA.80.041604} {\bibfield  {journal} {\bibinfo
  {journal} {Phys. Rev. A}\ }\textbf {\bibinfo {volume} {80}},\ \bibinfo
  {pages} {041604} (\bibinfo {year} {2009})}\BibitemShut {NoStop}%
\bibitem [{\citenamefont {Wang}\ \emph {et~al.}(2015)\citenamefont {Wang},
  \citenamefont {Liu}, \citenamefont {Iazzi}, \citenamefont {Troyer},\ and\
  \citenamefont {Harcos}}]{Wang2015}%
  \BibitemOpen
  \bibfield  {author} {\bibinfo {author} {\bibfnamefont {L.}~\bibnamefont
  {Wang}}, \bibinfo {author} {\bibfnamefont {Y.-H.}\ \bibnamefont {Liu}},
  \bibinfo {author} {\bibfnamefont {M.}~\bibnamefont {Iazzi}}, \bibinfo
  {author} {\bibfnamefont {M.}~\bibnamefont {Troyer}}, \ and\ \bibinfo {author}
  {\bibfnamefont {G.}~\bibnamefont {Harcos}},\ }\href {\doibase
  10.1103/PhysRevLett.115.250601} {\bibfield  {journal} {\bibinfo  {journal}
  {Phys. Rev. Lett.}\ }\textbf {\bibinfo {volume} {115}},\ \bibinfo {pages}
  {250601} (\bibinfo {year} {2015})}\BibitemShut {NoStop}%
\bibitem [{\citenamefont {Gorelik}\ \emph {et~al.}(2010)\citenamefont
  {Gorelik}, \citenamefont {Titvinidze}, \citenamefont {Hofstetter},
  \citenamefont {Snoek},\ and\ \citenamefont {Bl\"umer}}]{Gorelik2010}%
  \BibitemOpen
  \bibfield  {author} {\bibinfo {author} {\bibfnamefont {E.~V.}\ \bibnamefont
  {Gorelik}}, \bibinfo {author} {\bibfnamefont {I.}~\bibnamefont {Titvinidze}},
  \bibinfo {author} {\bibfnamefont {W.}~\bibnamefont {Hofstetter}}, \bibinfo
  {author} {\bibfnamefont {M.}~\bibnamefont {Snoek}}, \ and\ \bibinfo {author}
  {\bibfnamefont {N.}~\bibnamefont {Bl\"umer}},\ }\href {\doibase
  10.1103/PhysRevLett.105.065301} {\bibfield  {journal} {\bibinfo  {journal}
  {Phys. Rev. Lett.}\ }\textbf {\bibinfo {volume} {105}},\ \bibinfo {pages}
  {065301} (\bibinfo {year} {2010})}\BibitemShut {NoStop}%
\bibitem [{\citenamefont {Tang}\ \emph {et~al.}(2013)\citenamefont {Tang},
  \citenamefont {Paiva}, \citenamefont {Khatami},\ and\ \citenamefont
  {Rigol}}]{Tang2013}%
  \BibitemOpen
  \bibfield  {author} {\bibinfo {author} {\bibfnamefont {B.}~\bibnamefont
  {Tang}}, \bibinfo {author} {\bibfnamefont {T.}~\bibnamefont {Paiva}},
  \bibinfo {author} {\bibfnamefont {E.}~\bibnamefont {Khatami}}, \ and\
  \bibinfo {author} {\bibfnamefont {M.}~\bibnamefont {Rigol}},\ }\href
  {\doibase 10.1103/PhysRevB.88.125127} {\bibfield  {journal} {\bibinfo
  {journal} {Phys. Rev. B}\ }\textbf {\bibinfo {volume} {88}},\ \bibinfo
  {pages} {125127} (\bibinfo {year} {2013})}\BibitemShut {NoStop}%
\bibitem [{\citenamefont {Parsons}\ \emph {et~al.}(2016)\citenamefont
  {Parsons}, \citenamefont {Mazurenko}, \citenamefont {Chiu}, \citenamefont
  {Ji}, \citenamefont {Greif},\ and\ \citenamefont
  {Greiner}}]{parsons2016site}%
  \BibitemOpen
  \bibfield  {author} {\bibinfo {author} {\bibfnamefont {M.~F.}\ \bibnamefont
  {Parsons}}, \bibinfo {author} {\bibfnamefont {A.}~\bibnamefont {Mazurenko}},
  \bibinfo {author} {\bibfnamefont {C.~S.}\ \bibnamefont {Chiu}}, \bibinfo
  {author} {\bibfnamefont {G.}~\bibnamefont {Ji}}, \bibinfo {author}
  {\bibfnamefont {D.}~\bibnamefont {Greif}}, \ and\ \bibinfo {author}
  {\bibfnamefont {M.}~\bibnamefont {Greiner}},\ }\href {\doibase
  10.1126/science.aag1430} {\bibfield  {journal} {\bibinfo  {journal}
  {Science}\ }\textbf {\bibinfo {volume} {353}},\ \bibinfo {pages} {1253}
  (\bibinfo {year} {2016})}\BibitemShut {NoStop}%
\bibitem [{\citenamefont {Cheuk}\ \emph {et~al.}(2016)\citenamefont {Cheuk},
  \citenamefont {Nichols}, \citenamefont {Lawrence}, \citenamefont {Okan},
  \citenamefont {Zhang}, \citenamefont {Khatami}, \citenamefont {Trivedi},
  \citenamefont {Paiva}, \citenamefont {Rigol},\ and\ \citenamefont
  {Zwierlein}}]{cheuk2016observation}%
  \BibitemOpen
  \bibfield  {author} {\bibinfo {author} {\bibfnamefont {L.~W.}\ \bibnamefont
  {Cheuk}}, \bibinfo {author} {\bibfnamefont {M.~A.}\ \bibnamefont {Nichols}},
  \bibinfo {author} {\bibfnamefont {K.~R.}\ \bibnamefont {Lawrence}}, \bibinfo
  {author} {\bibfnamefont {M.}~\bibnamefont {Okan}}, \bibinfo {author}
  {\bibfnamefont {H.}~\bibnamefont {Zhang}}, \bibinfo {author} {\bibfnamefont
  {E.}~\bibnamefont {Khatami}}, \bibinfo {author} {\bibfnamefont
  {N.}~\bibnamefont {Trivedi}}, \bibinfo {author} {\bibfnamefont
  {T.}~\bibnamefont {Paiva}}, \bibinfo {author} {\bibfnamefont
  {M.}~\bibnamefont {Rigol}}, \ and\ \bibinfo {author} {\bibfnamefont {M.~W.}\
  \bibnamefont {Zwierlein}},\ }\href {\doibase 10.1126/science.aag3349}
  {\bibfield  {journal} {\bibinfo  {journal} {Science}\ }\textbf {\bibinfo
  {volume} {353}},\ \bibinfo {pages} {1260} (\bibinfo {year}
  {2016})}\BibitemShut {NoStop}%
\bibitem [{\citenamefont {Boll}\ \emph {et~al.}(2016)\citenamefont {Boll},
  \citenamefont {Hilker}, \citenamefont {Salomon}, \citenamefont {Omran},
  \citenamefont {Nespolo}, \citenamefont {Pollet}, \citenamefont {Bloch},\ and\
  \citenamefont {Gross}}]{boll2016spin}%
  \BibitemOpen
  \bibfield  {author} {\bibinfo {author} {\bibfnamefont {M.}~\bibnamefont
  {Boll}}, \bibinfo {author} {\bibfnamefont {T.~A.}\ \bibnamefont {Hilker}},
  \bibinfo {author} {\bibfnamefont {G.}~\bibnamefont {Salomon}}, \bibinfo
  {author} {\bibfnamefont {A.}~\bibnamefont {Omran}}, \bibinfo {author}
  {\bibfnamefont {J.}~\bibnamefont {Nespolo}}, \bibinfo {author} {\bibfnamefont
  {L.}~\bibnamefont {Pollet}}, \bibinfo {author} {\bibfnamefont
  {I.}~\bibnamefont {Bloch}}, \ and\ \bibinfo {author} {\bibfnamefont
  {C.}~\bibnamefont {Gross}},\ }\href {\doibase 10.1126/science.aag1635}
  {\bibfield  {journal} {\bibinfo  {journal} {Science}\ }\textbf {\bibinfo
  {volume} {353}},\ \bibinfo {pages} {1257} (\bibinfo {year}
  {2016})}\BibitemShut {NoStop}%
\bibitem [{\citenamefont {Molina}\ \emph {et~al.}(2009)\citenamefont {Molina},
  \citenamefont {Dukelsky},\ and\ \citenamefont {Schmitteckert}}]{Molina2009}%
  \BibitemOpen
  \bibfield  {author} {\bibinfo {author} {\bibfnamefont {R.~A.}\ \bibnamefont
  {Molina}}, \bibinfo {author} {\bibfnamefont {J.}~\bibnamefont {Dukelsky}}, \
  and\ \bibinfo {author} {\bibfnamefont {P.}~\bibnamefont {Schmitteckert}},\
  }\href {\doibase 10.1103/PhysRevA.80.013616} {\bibfield  {journal} {\bibinfo
  {journal} {Phys. Rev. A}\ }\textbf {\bibinfo {volume} {80}},\ \bibinfo
  {pages} {013616} (\bibinfo {year} {2009})}\BibitemShut {NoStop}%
\bibitem [{\citenamefont {Bloch}\ \emph {et~al.}(2008)\citenamefont {Bloch},
  \citenamefont {Dalibard},\ and\ \citenamefont {Zwerger}}]{Bloch2008}%
  \BibitemOpen
  \bibfield  {author} {\bibinfo {author} {\bibfnamefont {I.}~\bibnamefont
  {Bloch}}, \bibinfo {author} {\bibfnamefont {J.}~\bibnamefont {Dalibard}}, \
  and\ \bibinfo {author} {\bibfnamefont {W.}~\bibnamefont {Zwerger}},\ }\href
  {\doibase 10.1103/RevModPhys.80.885} {\bibfield  {journal} {\bibinfo
  {journal} {Rev. Mod. Phys.}\ }\textbf {\bibinfo {volume} {80}},\ \bibinfo
  {pages} {885} (\bibinfo {year} {2008})}\BibitemShut {NoStop}%
\end{thebibliography}
%

\end{document}